\documentclass[prc,aps,twocolumn]{revtex4}
\usepackage{graphicx}

\begin{document}
\date{\today}


\title{Isospin Transport at Fermi Energies}

\author
{ V.Baran$^{1,2}$, M.Colonna$^1$, 
M.Di Toro$^1$, M.Zielinska-Pfab$\acute{e} ^{3}$ 
and H.H.Wolter$^4$}
\affiliation{$^{1}$ Laboratori Nazionali del Sud, Via S. Sofia 44,
I-95123 Catania, Italy\\
and Physics-Astronomy Dept., University of Catania}

\affiliation{$^{2}$ NIPNE-HH, Bucharest and Bucharest University, Romania}

\affiliation{$^{3}$ Smith College, Northampton, MA 01063, USA}

\affiliation{$^{4}$ Dept. Physik, University of Munich, 85748 Garching, Germany}


\begin{abstract}
In this paper we investigate isospin transport mechanisms
in semi-peripheral collisions at Fermi energies. 
The effects of the formation of
a low density region (neck) between the two reaction partners 
and of pre-equilibrium emission
on the dynamics of isospin equilibration are carefully 
analyzed. 
We clearly identify two main contributions to the isospin 
transport: isospin diffusion due to the $N/Z$ ratio
and isospin drift due to the density gradients. Both effects are 
sensitive to the symmetry part of the nuclear Equation of State (EOS),
in particular to the value and slope around saturation density.   
\end{abstract}
\maketitle

keywords:
Isospin transport; Binary collisions; Neck Dynamics; Symmetry energy. \\ 
PACS numbers: 21.30.Fe, 25.70.-z, 25.70.Lm, 25.70.Pq. 

\section{INTRODUCTION}

In the last few years the increased accuracy of the experimental
techniques has renewed interest in nuclear reactions
at Fermi energies. Exclusive measurements, event-by-event analysis,
and a $4\pi$ coverage allow a deeper investigation of the evolution
of the reaction mechanisms with beam energy and centrality. New
insights into the understanding of the nuclear matter equation-of-state ($EOS$)
were gained \cite{DanielewiczSC298}.
In particular, recent experimental and theoretical analyses
were devoted to the study of the properties and
effects of the symmetry term of the $EOS$ (asy-$EOS$)
away from saturation conditions \cite{BaoAnBook01,baranRP}.

Indeed, the two-component character of nuclear matter
adds some special interest to the dynamics of 
heavy ion collisions at intermediate energies,
between $20$ and $100 AMeV$. In {\it central} collisions
isospin distillation is an important effect in multifragmentation
of charge asymmetric systems. Here phase separation 
is driven by isoscalar-like unstable
fluctuations, i.e. local in phase variations of
proton and neutron densities
\cite{BaranPRL86,ColonnaPRL88,BaranNPA703,Jer}.
This leads to a more symmetric ``liquid'' phase of fragments
surrounded by a more neutron rich ``gas'' relative to the 
original asymmetry of the system. Isoscaling phenomena,
observed experimentally, provide indications for
such a scenario \cite{XuPRL85,geraci}.

In {\it semi-peripheral} collisions between nuclei
with different $N/Z$ ratio, isospin dynamics 
will drive the system toward a uniform asymmetry
distribution. The degree of equilibration, correlated to the
interaction time, should provide some insights into transport
properties of fermionic systems
\cite{UehlingPR43,HellundPR56}, in particular give information 
on transport coefficients of asymmetric nuclear matter
\cite{AndersonPRB35,ShiPRC68}.

The aim of this work is to investigate the isospin
transfer through the neck region 
in semi-peripheral collisions of asymmetric nuclei
at Fermi energies.
Of particular interest is the role of the density 
dependence of the symmetry
energy in this process. 
The isospin transfer was measured for collisions of different
 $Sn$ isotopes at MSU  \cite{TsangNPA734,TsangPRL92,note1} and interpreted
 theoretically with the result that the asy-$EOS$ 
 should be rather stiff. 
 In these works the effect of pre-equilibrium emission, 
 which changes the isospin content of the interacting system,
  was not analyzed explicitly. Here we will discuss 
 this questions in detail, as well as the different transport 
 processes affecting the final isospin content.
 Quantitatively, dynamical isospin effects
can be properly understood only from microscopic calculations
based on transport models. We will base our study
on a stochastic BNV transport model (see refs. \cite{ColonnaNPA642,ColonnaRep}
for more details on the main ingredients of this approach).

\section{ISOSPIN EQUILIBRATION PROCESS}

We are focusing on the charge asymmetric collision
$^{124}Sn+^{112}Sn$, at $50 AMeV$ bombarding energy,
to which we refer as the mixed system, $(M)$.
To investigate the density ($\rho$) dependence we consider here two 
representative parameterizations of 
the symmetry energy, $E_{sym}(\rho,I)/A = C_{sym}(\rho)I^2, I = (N-Z)/A$ : 
one showing a rapidly increasing behaviour 
 with density, roughly proportional to $\rho^2$ (asysuperstiff)
and one where a kind of saturation is observed above normal
density (soft, $SKM^*$) (see Ref.\cite{baranRP,BaranNPA703} for more detail).

The BNV simulations have been performed for semi-peripheral collisions
at impact parameters $b = 6,8,9,10 fm$. In the last two cases
the reaction has dominantly a binary character and
the charge asymmetry of primary
projectile (target)- like fragments,
$PLF$ ( $TLF$), should provide the essential information
about the isospin equilibration rate.
At $b=8fm$ already about $25 \%$ of the events
are ternary. An IMF can be formed in the mid-velocity
region by neck fragmentation \cite{BaranNPA730}.
For more central events this mechanism
becomes dominant: at $b=6fm$, one or two IMF's are found in    
more than $70 \%$ of events \cite{BaranNPA703}. The fragment formation
in the neck region will influence the final isospin distribution
of the PLF/TLF,
and will render considerably more difficult the interpretation 
of the results. Thus
 in the following
we select only binary events in our analysis. 
We define the average interaction time, $t_c$, as the time elapsed
between the initial touching and the
moment when  $PLF$ and  $TLF$
reseparate. From our simulations we obtain
$t_c \approx 140, 120, 100, 80 fm/c$ for the  
impact parameters $b= 6, 8, 9, 10fm$, respectively.
Four hundred events were calculated for each initial
condition and for each asy-$EOS$.

\begin{figure}
\centering
\includegraphics*[scale=0.42]{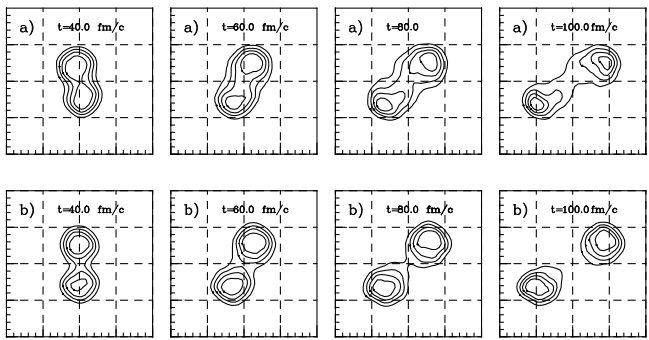}
\caption{\it $^{124}Sn+^{112}Sn$ $b=8fm$ (upper row) and 
$b=10fm$ (lower row) collision: density contour plots. The side of
each box is 40 fm.}
\label{difusden810}
\end{figure}

Typical density contour plots, at $b=8 fm$ and $b=10 fm$, 
are shown in Figure \ref{difusden810}. We note
the dynamical evolution of the overlap region:
driven by the
fast leading motion of $PL-$ and $TL-$ prefragments,
the formation of a lower density interface can be clearly
observed after around $40 fm/c$. An isospin migration, 
or transport, takes place during
this transient configuration of two residues
with densities close to the normal one, separated by a dilute
neck region. In contrast, in deep-inelastic collisions
at lower energies the isospin equilibration is driven
by the N/Z difference between the interacting nuclei 
having a quite uniform density profile 
without a low density interface until separation
\cite{FarineZPA339}.

We quantify the degree of equilibration by
the isospin transport (imbalance) ratio \cite{RamiPRL84}, defined as:

\begin{equation}
R_{i} = \frac{ 2 I_{i}^{M} - I_{i}^{H} - I_{i}^{L}}
{I_{i}^{H} - I_{i}^{L}}
\label{imbal}
\end{equation}

Here $i=P,T$ stands for the projectile-like (target-like) fragment. 
The quantities $I_i$
refer to the isospin, $I=(N-Z)/A$, or in general to any
isospin dependent quantity, characterizing the fragments 
at separation time,
for the mixed reaction $(M, 124+112)$,
the reactions between neutron rich ($H, 124+124$), 
and between neutron poor nuclei ($L, 112+112$), respectively.
A value of $R_{i}$ approaching zero is an indication of
a larger degree of equilibration. The extreme cases
$R_{T}= -1$ and $R_{P}= 1$ correspond to the absence of any
isospin transfer. 

\begin{figure}
\centering
\includegraphics*[scale=0.7]{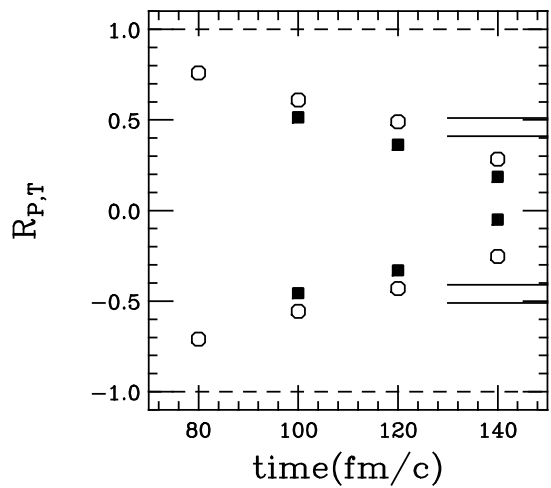}
\caption{\it 
The isospin transport ratio $R_{P,T}$, eq. (1), for the asysoft (full squares)
and asysuperstiff (circles) $EOS$'s as a function of the interaction 
time $t_c$, corresponding to different impact parameter $b$. 
The band between the two solid lines
corresponds to the experimental data of \cite{TsangPRL92,note1}. 
}
\label{imbalance3}
\end{figure}

The results of the calculations are shown in Figure \ref{imbalance3}
for the dependence of $R_{P/T}$ on the interaction time $t_c$
 for the asysoft (squares)
and asysuperstiff (circles) $EOS$'s. The figure also shows the
experimental values extracted in ref. \cite{TsangPRL92,note1} 
at semi-peripheral collisions at about $b=8 fm$.
We conclude, as in ref. \cite{TsangPRL92}, that 
an asystiff-like
$EOS$ provides a better agreement with the experimental
observations.

A significant difference between the two equations of state
is evident for larger interaction times, i.e. smaller values
of the impact parameter, $b=6,8fm$. The smaller values
of isospin transport ratios for the asysoft $EOS$
point toward a faster equilibration rate.
In refs. \cite{TsangPRL92,Chen04} an explanation was
based on the observation that below normal
density the asysoft $EOS$ has a larger value of the symmetry
energy. Therefore an enhanced isospin equilibration
will occur if the diffusion takes place
at uniform lower density. We intend to show, 
that in fact the mechanism
of charge equilibration is more complicated
due to dynamical evolution of the reaction at these energies.
Fast particle emission and density gradients will also play a role.
In the next section we
investigate more in detail the various
influences on the isospin transfer process. 

\section{ISOSPIN SHARING AT FERMI ENERGIES}

The isospin content of the two residues in a 
mixed collision system at separation time
is determined by the interplay between the
particle emission to the gas from each nucleus during the overlap
and the transfer of nucleons through the neck. We thus write 
simple balance equations:

\begin{eqnarray}
I_P = \frac{A_P^0}{A_P}( I_{P}^0 - \frac{A_{gP}}{A_{P}^0} I_{gP} -
\frac{A_{PT}}{A_{P}^0} I_{PT} + \frac{A_{TP}}{A_{P}^0} I_{TP})
\label{ipm} \\
I_{T} = \frac{A_{T}^0}{A_T}( I_{T}^0 - \frac{A_{gT}}{A_{T}^0}I_{gT} +
\frac{A_{PT}}{A_{T}^0}I_{PT} -\frac{A_{TP}}{A_{T}^0}I_{TP}) 
\label{itm}
\end{eqnarray}

Here $I_{P}$ ($I_{T}$) and $A_{P}$ ($A_{T}$)
are the $PLF$ ($TLF$) asymmetry and mass at separation, 
$I_{P}^0$ ($I_{T}^0$) and  $A_{P}^0$ ($A_{T}^0$)
the initial projectile (target) asymmetry and mass.
Then $I_{gP}$ ($I_{gT}$), $A_{gP}$ ($A_{gT}$) are the
asymmetries and masses of the projectile/target ``gas'', 
i.e. of the pre-equilibrium
particles emitted by the projectile (target) during the interaction time.
Finally $I_{PT}$ ($I_{TP}$), and $A_{PT}$ ($A_{TP}$)
are the asymmetry and mass of all nucleons transferred
from projectile (target) to target (projectile).

\begin{figure}
\centering
\includegraphics*[scale=0.45]{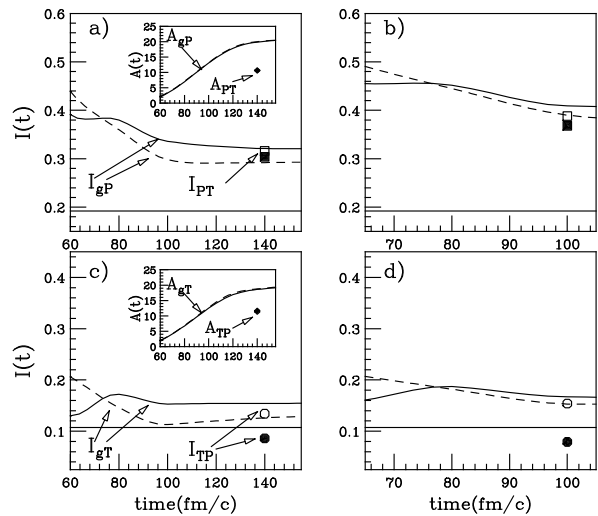}
\caption{\it $^{124}Sn+^{112}Sn$ collision at $b=8fm$ (left) and 
$b=10fm$ (right): Time
evolution of isospin content and mass of 
preequilibrium particles $I_{gP}$, $A_{gP}$ 
($I_{gT}$, $A_{gT}$) 
emitted from the projectile (top panels)
(resp. target, lower panels) 
for asysoft
(solid lines) and asysuperstiff (dashed lines) $EOS$.
The squares (circles) indicate the mass and   
isospin of the nucleons that have migrated from
projectile to target $I_{PT}$ (and vice versa, $I_{TP}$) for
asysoft (full symbols) and asysuperstiff (empty symbols)
at separation time.  The horizontal lines in the left panels give 
the initial isospin of the projectile, $I_P^0$, and target,$I_T^0$,
respectively.}
\label{agasiso}
\end{figure}

\begin{figure}
\centering
\includegraphics*[scale=0.45]{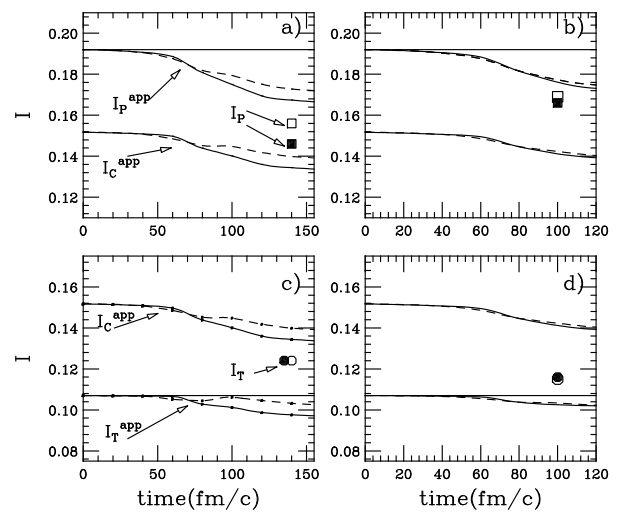}
\caption{\it $^{124}Sn+^{112}Sn$  at $b=8fm$ (left windows) and
$b=10fm$ (right windows):
Influence of pre-equilibrium emission on isospin content of
projectile (top) and target (bottom) systems. The curves show 
the ``apparent'' 
isospin without transfer, while the symbols denote the real 
isospin at separation time including also tranfer. 
Dashed lines and open symbols
correspond to asysuperstiff $EOS$ and the solid lines
and full symbols to the asysoft case. }
\label{isoevol}
\end{figure}

In Figure \ref{agasiso} we plot the time evolution of the
quantities $I_{gP}$, $I_{gT}$ and $A_{gP}$, $A_{gT}$,
as well as the values of $I_{PT}$  and $I_{TP}$
for the asysoft and the asysuperstiff $EOS$ for
two impact parameters. We note that $I_{gP}$ 
is much larger than $I_{P}^0$. The same is true for the
target but the difference is smaller. Thus
the pre-equilibrium emission reduces the $N/Z$ difference
between the two nuclei, competing with the transfer process.
This is also clearly seen in Figure \ref{isoevol} where
the time evolution of the projectile, target, and  
composite system apparent asymmetry,
$I^{app}_P$, $I^{app}_T$ and  $I^{app}_C$ is shown. The apparent
asymmetry is
calculated taking into account only the 
preequilibrium emission but not the isospin transfer inside the matter.
A stronger variation is observed
for the neutron rich projectile. The composite system asymmetry, 
$I_{C}^{app}$, 
is the limiting value toward which the two participants would evolve
in complete isospin equilibrium. In contrast, we show in the 
figure the final asymmetries $I_{P}$ and $I_{T}$ 
including the transfer processes.

Comparing the results for the two $EOS$'s in Figs. 3 and 4 we see that
a more neutron rich 
composition of pre-equilibrium emission is generated in the asysoft case 
because below normal density, from where most of the emitted nucleons
 originate, the neutrons (protons) are
less (more) bound than for the asysuperstiff $EOS$. 
The differences between the two asy-$EOS$'s are strongly
reduced at larger impact parameters, as seen in the results
for $b=10fm$ in Figures \ref{agasiso} and \ref{isoevol},
since the interaction times are much shorter.

\begin{figure}
\centering
\includegraphics*[scale=0.45]{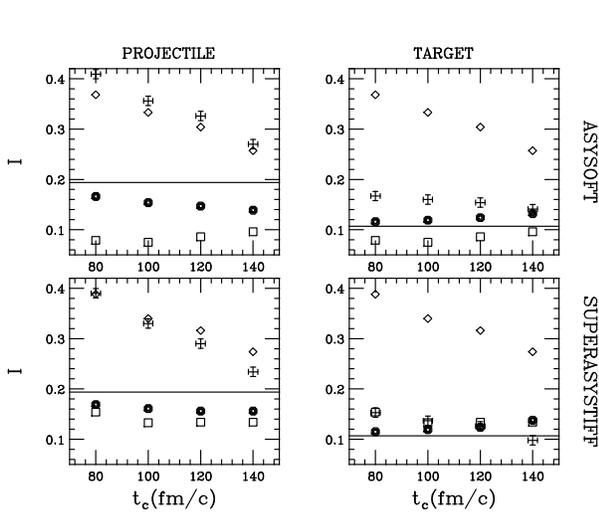}
\caption{\it $^{124}Sn+^{112}Sn$ collision with
 asysoft (top) and asysuperstiff (bottom) $EOS$
 as a function of collision time or impact parameter. 
 Left panels:
 $I_{P}$ (black circles), $I_{gP}$ (crosses),
$I_{PT}$ (rhombs) and $I_{TP}$ (squares).
Right windows: $I_{T}$  (black circles), $I_{gT}$ (crosses),
$I_{PT}$ (rhombs) and $I_{TP}$ (squares).
The horizontal solid lines refer to the
initial asymmetry of the projectile (target).}
\label{difusionpap3}
\end{figure}

We next show in Figure \ref{difusionpap3} the dependence 
of the quantities introduced above on 
interaction time, i.e. impact parameter. 
Here the projectile (target) and gas
final asymmetries $I_{P}$ ,$I_{gP}$ ($I_{T}$, $I_{gT}$)
are confronted with the transferred asymmetries 
$I_{PT}$ and $I_{TP}$. 
For the projectile, that is neutron rich, both pre-equilibrium emission and nucleon
transfer drive the system toward a more
symmetric configuration (left). The two processes have 
opposite effects and thus tend
to compensate for the target (right). Therefore the projectile 
asymmetry has a more pronounced deviation
from the corresponding initial value in comparison to the target.  

We find a a clear dependence on the
asy-$EOS$ of the isospin transferred between the
two nuclei. In particular, for $b=8fm (t_c = 120 fm/c)$ we observe that
\begin{eqnarray}
I_{PT}^{(asysuperstiff)} > I_{PT}^{(asysoft)} > I_{P}^{0}=0.192 \\
I_{TP}^{(asysuperstiff)} > I_{T}^{0}=0.107 > I_{TP}^{(asysoft)} \label{inequal}
\end{eqnarray}
We will show in the next chapter
 that the origin of these inequalities lies
in the existence of the low density interface
and the density dependence of symmetry energy.
The asysuperstiff $EOS$ favors the neutron migration
toward the neck region from both participants. This
explains why simultaneously $I_{PT}^{(asysuperstiff)} > I_{P}^{0}$ and
$I_{TP}^{(asysuperstiff)} > I_{T}^{0}$.
We will see that for the asysoft $EOS$ this effect is weakened.

\section{INTERPRETATION OF THE RESULTS}

The above arguments can be made more explicit by considering that
the proton and neutron migration is
dictated by the spatial gradients of
the corresponding chemical potentials $\mu_{p/n}(\rho_p,\rho_n,T)$
 \cite{balian}.
The currents of the two species can be expressed as follows 
\begin{eqnarray}
j_{n} = - ct \nabla \mu_{n}(\rho_p,\rho_n,T) = - ct [
\left({\partial \mu_n \over \partial \rho_n}\right)_{\rho_p,T} \nabla  \rho_n + \nonumber  \\
\left({\partial \mu_n \over \partial \rho_p}\right)_{\rho_n,T} \nabla  \rho_p ] \nonumber  \\
j_{p} = - ct \nabla \mu_{p}(\rho_p,\rho_n,T) = - ct [
\left({\partial \mu_p \over \partial \rho_n}\right)_{\rho_p,T} \nabla  \rho_n + \nonumber  \\ 
\left({\partial \mu_p \over \partial \rho_p}\right)_{\rho_n,T} \nabla  \rho_p ] \nonumber,
\end{eqnarray}
where $ct$ is a constant.
Rewriting these expressions in terms of Landau parameters, 
 $F_0^{qq'}$ (\cite{baranRP} and references therein), using:  
\begin{equation}
N_q(T){\partial\mu_q \over \partial\rho_{q'}}
= \delta_{qq'} + F_0^{qq'},~~~~~~~q=n,p~~~~~~q'=n,p    \label{muF0}
\end{equation}
(where $N_q(T)$ is the level density),
we obtain:
\begin{eqnarray}
j_{n} &=& -\frac{ct}{2} ([ N_n^{-1}(1+I) + f^{\rho}_n 
+  I f^{I}_n )] \nabla  \rho + 
 \rho [ N_n^{-1} + f^{I}_n ] \nabla I) \nonumber  \\ 
&=& -D_{n}^{\rho} \nabla  \rho - D_{n}^{I} \nabla I \\
j_{p} &=& -\frac{ct}{2} ([ N_p^{-1}(1-I) + f^{\rho}_p 
-  I f^{I}_p )] \nabla  \rho - 
 \rho [ N_p^{-1} + f^{I}_p ] \nabla I) \nonumber  \\ 
&=& -D_{p}^{\rho} \nabla  \rho - D_{p}^{I} \nabla I, 
\end{eqnarray}
where
\begin{eqnarray}
f^{\rho}_q & = &  N_q^{-1}(F_0^{qq}+F_0^{qq'}) = \frac{\partial U_q}
{\partial \rho_q} + \frac{\partial U_q}{\partial\rho_{q'}} \label{frho}~~, \\
 f^{I}_q & = &  N_q^{-1}(F_0^{qq}-F_0^{qq'}) = \frac{\partial U_q}{\partial 
\rho_q} - \frac{\partial U_q}{\partial\rho_{q'}}~~;(q \neq q')~~. \label{fiso}
\end{eqnarray} 
$U_q$ is the neutron or proton mean-field potential.
The second lines in eqs. (7,8) define $p$ and $n$ drift and 
diffusion coefficients
due to density and isospin gradients,
 $D_{q}^{\rho}$ and $D_{q}^I$, respectively.
The terms $N_q^{-1}(1 \pm I)$ and $f^{\rho}_q \pm I f^{I}_q$, 
that appear in the
density drift coefficients $D_{q}^{\rho}$, can be expressed as
$$ N_q^{-1}(1 \pm I) = 2 N^{-1}(\rho,T) \pm 4I
\frac{\partial C_{sym}^{kin}}{\partial\rho} + O(I^2);$$
\begin{equation}
f^{\rho}_{q} \pm I f^{I}_{q} = {\it F}(\rho) \pm 4I
\frac{\partial C_{sym}^{pot}}{\partial\rho} + O(I^{2}), (+ n, - p),
\label{Drho}
\end{equation}
where the function ${\it F}(\rho)$ depends only on the isoscalar 
part of the interaction. 
We have denoted by $C_{sym}^{kin}$
and $C_{sym}^{pot}$ the kinetic and the potential part of the
symmetry energy coefficient $C_{sym}$, respectively.  
Combining eqs. (\ref{Drho}), one can see that 
the isovector part of
the nuclear interaction enters the coefficients  $D_{q}^{\rho}$
through the derivative of the total symmetry energy $C_{sym}$.   
On the other hand the isospin diffusion coefficients $D_{q}^{I}$ 
are proportional to the quantity
\begin{equation}
\rho ( N_q^{-1} + f^{I}_q)  = 4 [C_{sym} \pm I(\rho\frac{\partial C_{sym}}{\partial\rho}-
C_{sym})],
\end{equation}
and depend, in leading order, on the value of the symmetry 
energy coeffcient $C_{sym}$.

Various particular situations can be derived
from these relations. In symmetric nuclear matter $D_{n}^{I}= - D_{p}^{I}$ and 
 $D_{n}^{\rho}= D_{p}^{\rho}$.
In the absence of density gradients the proton
current will flow oppositely and equal in magnitude to the neutron current.
On the other hand, for density gradients only, in asymmetric
nuclear matter the proton and neutron currents may have the same direction 
but assume different values, inducing isospin gradients \cite{dist}. 
Such a situation can be encountered in
semi-pheripheral collisions between identical, charge asymmetric
nuclei with the formation of a dilute intermediate region.

\begin{figure}
\centering
\includegraphics*[scale=0.4]{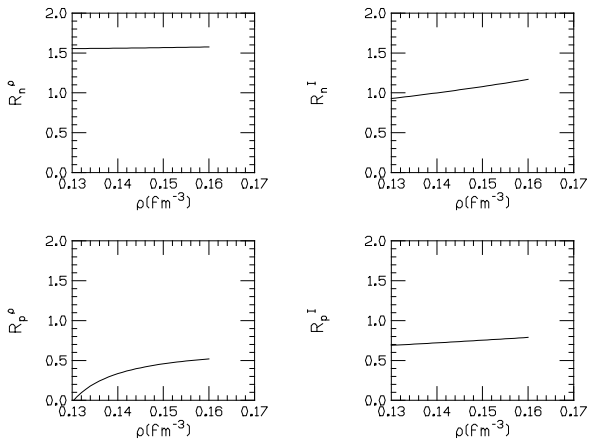}
\caption{\it Ratios of drift coefficients 
$R_q^i = D_{q}^{i, asysuperstiff}/D_{q}^{i, asysoft}$
,$i=\rho,I$ and $q=n,p$, as a function of the density for fixed asymmetry $I=0.2$.}
\label{difus1}
\end{figure}

For the two asy-$EOS$'s we calculate the
coefficients $D_{q}^i$, $i=\rho, I$ and $q=n,p$.
We plot the ratios 
$R_q^i = D_{q}^{i, asysuperstiff}  /  D_{q}^{i, asysoft}$
in Figure \ref{difus1}
as a function of the density for a fixed asymmetry $I=0.2$ .
These values of the asymmetry and density are close to the
physical conditions expected for the projectile or target region.
The only negative coefficient is $D_{p}^I$.
The isospin gradients, directed from the projectile to the neck and
from the neck to the target,
induce neutron and proton flows in opposite directions.
However the ratios of the corresponding
coefficients are quite close to unity for the two asy-$EOS$'s
and therefore the effects are similar.

Since the density gradient is oriented from projectile and target
residues to the neck, 
neutrons and protons migrate
from higher toward lower density regions. 
Around and below saturation density 
$R_n^{\rho} = D_{n}^{\rho,asysuperstiff}/D_{n}^{\rho,asysoft} > 1$ and 
$R_p^{\rho} = D_{p}^{\rho,asysuperstiff}/D_{p}^{\rho,asysoft} < 1$. These
inequalities suggest that more neutrons and less protons
migrate from projectile toward neck in the
case of asysuperstiff $EOS$ resulting in the formation of a more
neutron  rich intermediate region. This is due to 
the larger value of the derivative of the symmetry energy around
normal density and is in agreement with
the behavior observed in the numerical simulations, see eqs. (4,\ref{inequal}).

We note that for asysuperstiff $EOS$,
in spite of an enhanced isospin migration toward the neck at separation time,
the projectile residue is more asymmetric in comparison to the asysoft case.
One reason is that for the asysuperstiff $EOS$ during the
pre-equilibrium emission not as many neutrons are removed
as for the asysoft $EOS$. Also, as it was shown, for the asysuperstiff $EOS$ more asymmetric
matter is also transferred from the target.


We now estimate the effect of isospin transport and pre-equilibrium 
emission on the transport ratios. 
Within the following approximations 

\begin{eqnarray}
\frac{ A_{P}^{0} }{ A_{P} } {\Bigg|}_M \approx 
 \frac{ A_{P}^{0} }{ A_{P} } {\Bigg|}_H \approx
\frac{ A_{P}^{0} }{ A_{P} } {\Bigg|}_L ;
\frac{ A_{gP} }{ A_{P}^{0} }  {\Bigg|}_M \approx  
 \frac{ A_{gP} }{ A_{P}^{0} } {\Bigg|}_H
\approx \frac{  A_{gP} }{  A_{P}^{0} } {\Bigg|}_L  ;\\
A_{TP} \approx A_{PT} ; 
 I_{gP}^M \approx I_{gP}^{H}; I_{gT}^M \approx I_{gT}^L ~~,
\end{eqnarray}

where the indices $M,H,L$ refer again to the 
mixed (124+112), the neutron rich $(124+124)$ 
and the neutron deficient $(112+112)$ systems,
we arrive at a simplified expression for the isospin transport ratio
for the projectile 
which shows explicitely the dependence on the isospin transport 
$( I_{PT} - I_{TP} )$ and on the 
pre-equilibrium emission $( I_{gP}^{H} - I_{gP}^{L} )$:

\begin{equation}
R_{P} \approx 1 - \frac{ 2 \frac{A_{PT}}{ A_{P}^{0,H}} ( I_{PT} - I_{TP} )}
{ I_{P}^{0,H} - I_{P}^{0,L} - \frac{ A_{gP} }{ A_{P}^{0,H} }
( I_{gP}^{H} - I_{gP}^{L} ) }   
\end{equation}

With similar approximations 
the target isospin transport ratio can be expressed as: 
\begin{equation}
R_{T} \approx - 1 + \frac{2 \frac{A_{PT}} {A_{T}^{0,L}}( I_{PT} - I_{TP} )}
{I_{T}^{0,H} - I_{T}^{0,L} -\frac{A_{gT}}{A_{T}^{0,L}}(I_{gT}^{H} - I_{gT}^{L})}   
\end{equation}

It is observed that the transport ratios depend on
the difference $I_{PT} - I_{TP}$ as expected. 
However,  in contrast to what was assumed 
 in refs. \cite{TsangPRL92,Chen04},
 it is seen that they also depend on the pre-equilibrium emission 
 which reduces the absolute value of the transport ratios. 
Both effects are
smaller in the case of the asysuperstiff $EOS$, for which indeed
a larger $R$ ratio is obtained.

\section{CONCLUSIONS}

In this work we have studied processes related to isospin
equilibration in semipheripheral collisions at Fermi energies
and their dependence on the symmetry term of the $EOS$.
A special feature of these reactions
is the development of a low density interface between the two
residues. The neck region is controlling the proton
and neutron currents and their direction. The presence of
density gradients also affects the isospin exchange between
projectile and target and we have shown that this
is sensitive to the density dependence of the symmetry energy.
The neutron to proton ratio emitted during the interaction
stage is also influenced by the asy-$EOS$. The interplay between the two 
processes leads to a stronger equilibration for asy-soft $EOS$,
as it is evidenced by the isospin transport (imbalance) ratio.
Actually, in the asy-stiff case, a larger isospin transfer
is observed, due to the presence of density gradients, directed
from $PLF$ and $TLF$ towards the neck region. 
However, since we are studying binary processes, finally we observe a
kind of compensation between the asymmetry of the matter 
transferred from projectile to target ($I_{PT}$) and
from target to projectile($I_{TP}$).
>From this point of view, to put in better evidence effects due
to the presence of density gradients, it would be more appropriate 
to study events where fragments originating from the neck region
are also detected, possibly with their isospin content.  
 
In the present study a  rapidly increasing symmetry energy at subnormal densities 
(asysuperstiff) appears to be in better agreement with the
existing data \cite{TsangPRL92}. 
More recent work also considered momentum dependent interactions, 
showing that the more repulsive character of the overall
dynamics may reduce the symmetry energy stiffness required to reproduce
the data  \cite{Chen04}. 
The study of the interplay between the effects due to the isoscalar
and isovector part of the $EOS$ on isospin transport observables
deserves further attention.   

In conclusion, charge equilibration measurements in semi-peripheral
heavy ion collisions at Fermi energies provide new independent 
observables to study the 
poorly known density dependence of the symmetry term of the nuclear
EOS.  
This is of interest for other 
properties of asymmetric matter, like neutron skin and isovector
collective response in finite nuclei, and may also be important
for neutron star crust structures \cite{baranRP}.


\begin{thebibliography}{99}
\bibitem{note1}We note that the experimental impact parameter selection 
corresponds to rather peripheral collisions (estimated impact parameters
b$>$8 fm \cite{TsangPRL92}). In this sense the appropriate comparison should be with
$t_c$ = 110-120 fm/c interaction time simulations.
\bibitem{dist}We note that this is also the mechanism behind the isospin
distillation (n-enrichment of the gas phase) in the liquid-gas phase
transition for asymmetric matter \cite{baranRP,BaranPRL86}.
\end{thebibliography}
\end{document}